\documentclass[prb,twocolumn,superscriptaddress,amsmath,amssymb,aps,prb,longbibliography]{revtex4-2}
\usepackage[utf8]{inputenc}
\usepackage{amssymb}
\usepackage{amsmath}
\usepackage{braket}
\usepackage[english]{babel}
\usepackage{graphicx}

\usepackage{tikz}
\usepackage{graphicx,bm}
\usetikzlibrary{positioning}

\usepackage[T1]{fontenc}
\usepackage{lmodern}

\usepackage{indentfirst}
\usepackage{gensymb}
\usepackage{newunicodechar}
\newunicodechar{−}{\ensuremath{-}}
\usepackage{textcomp}
\begin{document}

\preprint{AIP/123-QED}

\title{Switching perpendicular magnets for Processing-in-memory with voltage gated Weyl Semimetals}
\author{Youjian Chen}
\affiliation{Department of Physics, University of Virginia, Charlottesville, Virginia 22904, USA}

\author{Hamed Vakili}
\affiliation{Department of Physics and Astronomy and Nebraska Center for Materials and Nanoscience, University of Nebraska, Lincoln, Nebraska 68588, USA}

\author{Md Golam Morshed}
\affiliation{Department of Electrical and Computer Engineering, University of Virginia, Charlottesville, Virginia 22904, USA}

\author{Avik W. Ghosh}
\affiliation{Department of Physics, University of Virginia, Charlottesville, Virginia 22904, USA}
\affiliation{Department of Electrical and Computer Engineering, University of Virginia, Charlottesville, Virginia 22904, USA}

\date{\today}
\begin{abstract}
Processing-in-memory (PIM) reduces data transfer latency by rolling memory and logic elements into one compute location. As an emergent material candidate for such an architecture, we propose a strained Weyl semimetal based spin–orbit-torque random-access memory (SWSM-SOTRAM) device. The spin–orbit torque (SOT) originates from two mechanisms: (1) the inverse spin Galvanic effect (iSGE), which generates nonequilibrium in-plane spin accumulation at interfaces, and (2) a bulk spin Hall effect (SHE), which produces a transverse spin current carrying out-of-plane spin angular momentum. The latter is tunable via an exchange Zeeman field. Both effects are evaluated using the tight-binding model coupled with a nonequilibrium Green’s function (TB-NEGF) formalism for quantum transport. Information write is achieved through SOT switching of an out-of-plane free magnet. A piezo attached to a magnetostrictive selector modulates the strain in the latter, leading to the rotation of the magnetization and hence the exchange Zeeman field exerted on the Weyl semimetal. This strain-controlled exchange field enables the symmetry tuning of the Weyl semimetal and modulation of its spin Hall effect. The TB-NEGF calculations of SHE and iSGE, combined with Landau–Lifshitz–Gilbert (LLG) simulations of magnetization dynamics, establish the SOT switching mechanism and demonstrate a pathway toward the SWSM-SOTRAM PIM device.
\end{abstract}

\maketitle
\section{\label{sec:intro}Introduction}

Processing-in-Memory (PIM), also called Compute-in-Memory, is an architecture that integrates both memory and logic elements into a single computing fabric \cite{chi2016prime,ghose2019processing}. 
Compared to traditional Harvard or Von Neumann architectures, there is no long-range movement of data between the memory and processing elements in a PIM, making this architecture faster and more energy-efficient \cite{fong2015spin}. A convenient PIM architecture is one with a set of vertically integrated structures that can combine memory and logic operations. In particular, for non-volatile magnetic memory, it is desirable to employ a channel material that is both spin selective and gate tunable. 

A recently proposed Strained-Topological-Insulator based Spin-Orbit-Torque Random-Access-Memory (STI-SOTRAM) bitcell \cite{morshed2025strained,vakili2022low} aims to be a scalable candidate for a PIM architecture. STI-SOTRAM consists of a multilayer stack that simultaneously accesses a magnetic memory element sitting above with in-plane anisotropy, and a logic driver with a strain gated out-of-plane magnet sitting below, with a single intervening quantum material that is both spin selective and gate tunable, in this case, a topological insulator (TI). When an external bias voltage is applied to a piezo below the gating/selector magnet, the stress-induced field rotates the latter between out-of-plane and in-plane \cite{roy2012energy,biswas2017experimental,fashami2011magnetization}, alternately gapping and activating the TI surface states. The activated TI generates currents consisting of in-plane spin angular momentum with a high charge-to-spin conversion coefficient, exerting a damping-like torque to rotate the storage magnet between $+\hat{y}$ and $-\hat{y}$ directions depending on the drain bias. Activating two rows of PIM bitcells in a crossbar geometry thereafter allows us to execute a logic operation (bitwise AND, OR, or MAC - Multiply-And-Accumulate for synaptic sums) using a sense amplifier with a tunable threshold. As selector and storage magnets are co-located in a vertical geometry, the structure is scalable and naturally suited for a PIM architecture that pre-processes stored data locally, within one vertically integrated, compact bit cell. 
Dynamic simulations of our STI-SOTRAM show potential energy-latency savings compared to conventional DRAM. However, since the TI surface can only support in-plane spins, the storage magnet also needs to be in-plane, which makes it less reliable and scalable than a perpendicular magnet. 

To successfully torque an out-of-plane magnetization between $+\hat{z}$ and $-\hat{z}$ directions, we need materials that deliver out-of-plane spin angular momentum while retaining gate tunability. 
 We argue that topological Weyl semimetals (WSM) can accomplish both of these tasks. With at least one of time-reversal or spatial-inversion symmetry broken, a Weyl semimetal can host pairs of Weyl fermions as emergent quasiparticles\cite{murakami2007phase,murakami2007tuning,yang2011quantum,burkov2011topological,burkov2011weyl,volovik2003universe,wan2011topological,xu2011chern}. 
 In contrast to a TI, where the inverse spin Galvanic effect (iSGE) accumulates a steady nonequilibrium in-plane spin density on its surface that can subsequently diffuse into a top magnetic layer and deliver a precessional torque \cite{manchon2019current,mellnik2014spin,fischer2016spin,han2017room,vakili2022low},  a WSM can generate  a sizable spin current modulated by its crystalline symmetries\cite{song2020coexistence,ozawa2024effective}. For example, WSM can generate 
 a transverse spin current carrying out-of-plane spin angular momentum via the bulk spin Hall effect (SHE) with an exchange (Zeeman) field. Such a  transverse spin current can then exert an anti-damping torque to switch the out-of-plane magnetization and write information into a free magnet with ±$\hat{z}$ magnetization.
 
 In this paper, we model Weyl semimetals and compute their SOT for out-of-plane magnetization switching. A lot of materials with $D_{2d}$ point group symmetry are predicted to be spatial-inversion symmetry breaking WSMs \cite{ruan2016ideal,ruan2016symmetry} , and can potentially carry out-of-plane spin angular momentum to switch a storage magnet between $+\hat{z}$ and $-\hat{z}$ directions. Breaking time-reversal and crystalline symmetry would create these transverse spin currents, with their strengths set by the ratio of iSGE to SHE and controlled by a voltage rotated selector magnet. These properties allow us to design a strained-Weyl semimetal spin-orbit torque random-access memory (SWSM-SOTRAM) for in-memory computing.

\section{\label{sec: Method}Method}

\subsection{\label{sec:WSMmodel}Model for Weyl semimetal layer}

\begin{figure}[h!]
\includegraphics[width=\columnwidth]{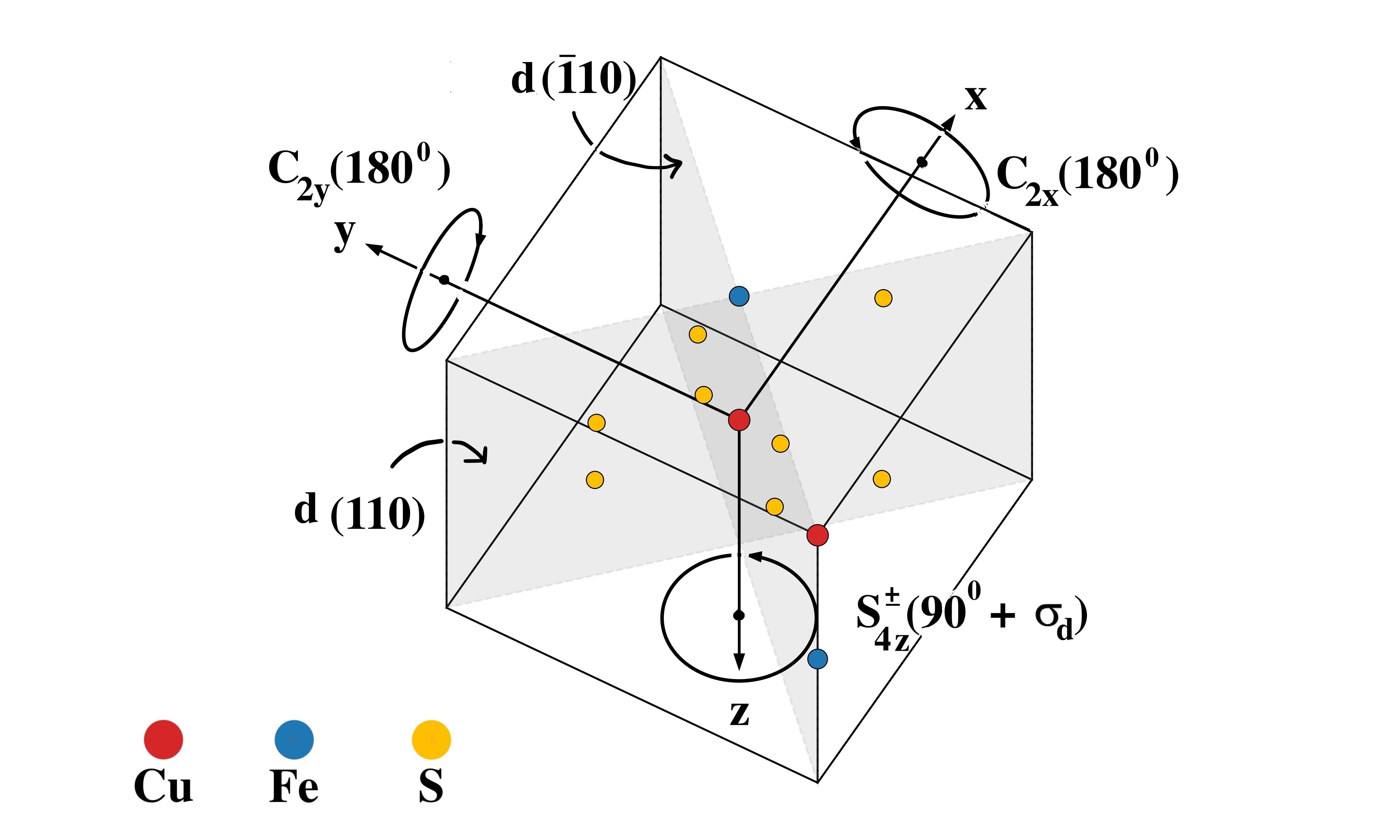}%
\caption{\label{fig:D2d}Example of a D$_{2d}$ unit cell, in this case CuFeS$_2$ (chalcopyrite, $I\bar{4}2d$). The space group contains: twofold screw-axis-two-fold rotations about $x$ and $y$ axes $C_{2x}$ and $C_{2y}$ combined with a translation of $1/2$ along each axis; S$_{4z}^{+}$ and S$^{-}_{4z}$, improper four-fold rotations about z-axis (90$^\circ$ + reflection); two-fold rotational symmetry $C_{2z}$-product of two improper rotation($C_{2z}=(S_{4z}^{+})^2$); glide mirror $\sigma_d (M_{xy})$ and $\sigma_d (M_{-xy})$ - dihedral mirrors  plus translation t = $(1/4, 1/4, 0)$ that contain the $z$-axis and bisect $x$ and $y$.}
\end{figure}

\begin{figure*}
\includegraphics[width=\textwidth]{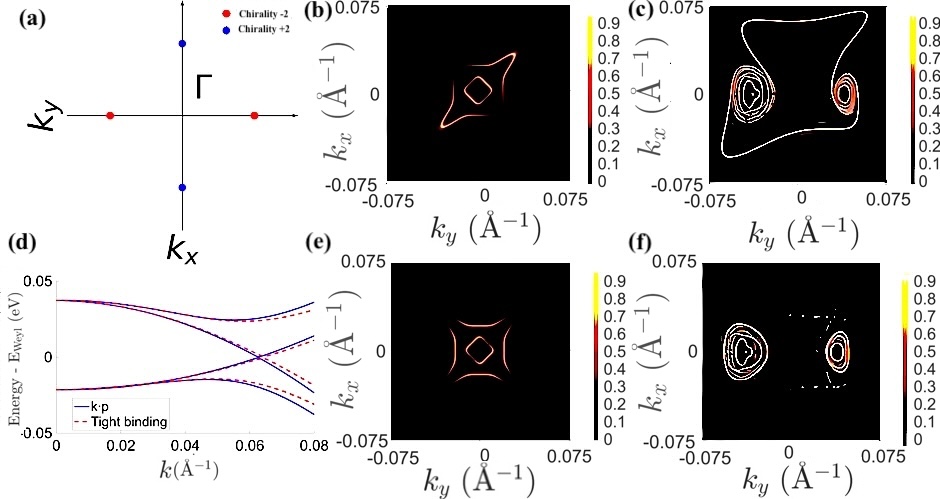}%
\caption{Spectrum of the four band model for Weyl semimetals with $D_{2d}$ point group symmetry.\label{fig:Band and Symmetries} Fermi-arc and band plots are computed using the four-band model of Ref.~\citenum{ruan2016ideal}. (a) projection of Weyl points in the $k_x-k_y$ plane. The chalcopyrite Weyl semimetal hosts eight symmetry-related Weyl points, grouped into two sets - $(\pm k_{x}^{*},0,\pm k_{z}^{*})$ with chirality $+1$ for each Weyl point and chirality $+2$ for projections in $k_x-k_y$ plane , and $(0,\pm k_{y}^{*},\pm k_{z}^{*})$ with chirality $-1$, where $k_{x}^{*}=k_{y}^{*}$ for each Weyl point and chirality $-2$ for projections in $k_x-k_y$ plane. (b) The top surface of a $50$-layer Weyl semimetal slab without an exchange (Zeeman) field, using low energy model of $\mathrm{CuTlTe_2}$ as an example \cite{ruan2016ideal}, showing Fermi arcs that connect the Weyl points. (c) The top surface with an exchange (Zeeman) field applied along $+\hat{y}$, of magnitude $\Delta_{\mathrm{ex}} = 0.025 ~\mathrm{eV}$, again showing Fermi arcs that connect the Weyl points. (d) Band structure along the $\Gamma$–Weyl point path, comparing the $\bm k\cdot\bm p$ and tight-binding models. The $k$ path follows the direction from $\Gamma = (0,0,0)$ to the Weyl point at $(k_{x}^{*}, 0, k_{z}^{*})$. (e) Bulk of a $50$-layer Weyl semimetal slab without an exchange (Zeeman) field, showing Fermi arcs that connect the Weyl points. (f) Bulk with an exchange (Zeeman) field applied along $+\hat{y}$, of magnitude $\Delta_{\mathrm{ex}} = 0.025 ~\mathrm{eV}$, again showing Fermi arcs that connect the Weyl points.
}
\end{figure*}

The key component of the PIM device is a Weyl semimetal that converts charge current into a transverse spin current carrying out-of-plane spin angular momentum. The specific class of materials that we explore for the PIM device is Weyl semimetals with $D_{2d}$ point group symmetry (Fig.~\ref{fig:D2d}), which includes a twofold rotation along $z$ ($C_{2z}$), two twofold rotations ($C_{2x}$ and $C_{2y}$), two diagonal mirrors ($M_{xy}$ and $M_{-xy}$), and two improper rotations ($S_{4z}^{+}$ and $S_{4z}^{-}$). In addition to point group symmetries, time-reversal symmetry ($\mathcal{T}$) is preserved, allowing these materials to have eight Weyl points, as shown in Fig.~\ref{fig:Band and Symmetries}(a). Examples of such Weyl semimetal materials include chalcopyrite compounds such as $\mathrm{CuTlTe_2}$ with space group $I\bar{4}2d$ \cite{ruan2016ideal}, strained-mercury chalcogenides such as strained-$\mathrm{HgTe}$, and strained-half-Heusler compounds such as strained-$\mathrm{LaPtBi}$ with space group $I\bar{4}2m$ \cite{ruan2016symmetry}.

 In the linear-$\cal{E}$ regime, the transverse spin current density with polarization $\sigma_k$ and direction $i$ could be expressed as $j^{\sigma_k}_{i}=\sigma^{k}_{ij}(\vec{B}){\cal{E}}_{j}$, where ${\cal{E}}_j$ is the $j$-th Cartesian component of the applied electric field and $\sigma^{k}_{ij}(\vec{B})$ is the exchange (Zeeman) field-dependent spin Hall conductance. We can expand the $\sigma^{k}_{ij}(\vec{B})$ in $\vec{B}$ to get $j^{\sigma_k}_i=\sum_{j}\sigma^{k}_{ij}{\cal{E}}_{j}+\sum_{jl}\alpha_{ijkl}{\cal{E}}_{j}B_{l}$, where ${\cal{E}}_j$ is the $j$-th Cartesian component of the applied electric field and $B_{l}$ is the $l$-th Cartesian component of the exchange (Zeeman) field. The spin Hall conductivity tensor $\sigma^{k}_{ij}=\sigma^{k}_{ij}(\vec{0})$ and the fourth-rank $TR$-even tensor $\alpha_{ijkl}=\partial \sigma^{k}_{ij}(\vec{B})/{\partial B_{l}}$ should respect the symmetries of group operations\cite{seemann2015symmetry,kleiner1966space}. In our paper, we have a net electric field along the $\pm \hat{y}$ direction and are primarily interested in the transverse spin current along the $\pm \hat{z}$ direction that carries out-of-plane spin angular momentum. Therefore, we are primarily focused on the spin Hall conductivity tensor $\sigma^{z}_{zy}$ and the fourth-rank tensor $\alpha_{zyzl}(l=x,y,z)$. The spin Hall conductivity tensor $\sigma^{z}_{zy}$ is symmetry-forbidden by rotational symmetries $C_{2x}$ and $C_{2z}$,each of which individually reverses its sign of $\sigma^{z}_{zy}$ , as well as by the combined action of the mirror symmetries $M_{xy}$ and $M_{-xy}$ and the combined action of improper rotation $S^{-}_{4z}$ and $S^{+}_{4z}$. For the fourth-rank tensor $\alpha_{zyzl}(l=x,y,z)$, the only nonzero component is $\alpha_{zyzy}=\alpha_{\textbf{planar}}$, where $\textbf{planar}$ stands for in-plane magnetoelectric couplings. In other words, the transverse spin current that carries out-of-plane spin angular momentum under a net electric field in the $\pm \hat{y}$ direction is
 \begin{equation}    J^{\sigma_z}_z({\cal{E}}_y,\vec{B})=\alpha_{\textbf{planar}} {\cal{E}}_y B_y.\label{eq:spin current1}
 \end{equation}
The bulk states of the Weyl semimetal can be described by a low energy four band $\vec{k}\cdot\vec{p}$ Hamiltonian in a spin-$\frac{3}{2}$ basis, up to quadratic order in momenta $\bm k$, as \cite{ruan2016ideal}:
 \begin{equation}
\begin{split}
H(\mathbf{k}) =\; & \epsilon_0(\mathbf{k})\mathbb{I}_{4\times 4} 
+ c_1 (k_y k_z \Gamma^1 + k_z k_x \Gamma^2) 
+ c_2 k_x k_y \Gamma^3 \\
&+ [c_3 (k_z^2 - m^2) + c_5 (k_x^2 + k_y^2)]\Gamma^5\\ 
&+ c_4 (k_x^2 - k_y^2) \Gamma^4 + v k_z \Gamma^{35} 
+ \alpha_1 (k_x \Gamma^{15} + k_y \Gamma^{25}) \\
&+ \alpha_2 (k_x \Gamma^{23} + k_y \Gamma^{13}) 
+ \alpha_3 (k_x \Gamma^{14} - k_y \Gamma^{24}).
\end{split}
\label{eq:kp}
\end{equation}
where $\epsilon_0(\mathbf{k}) = a_0 + a_1(k_x^2 + k_y^2) + a_2 k_z^2$.
The lattice constant $c = 12.8 ~\mathring{A}$ is about twice that of $a = 6.7 ~\mathring{A}$.
Here, $a_i$, $c_i$, and $m$ are coefficients, and $\Gamma^{a}$ denotes the Gamma matrices listed in Ref.~\citenum{ruan2016ideal}.

Based on Eq.~\eqref{eq:spin current1}, to engineer the symmetries and control the transverse spin current in the  Weyl semimetal, we consider a Weyl semimetal coupled to a selector magnet with magnetization along the $\hat{n}$ direction. Among the various effects induced by the ferromagnet–Weyl semimetal coupling, we retain only the lowest-order term, namely the exchange (Zeeman) field, defined as  
\begin{equation}
H_Z = \Delta_{ex}\ \hat{n} \cdot  \mathcal{\bm J}
\label{eq:zeeman}
\end{equation}
where $\hat{n}=(\sin\theta \cos\phi,
\sin\theta \sin\phi,
\cos\theta)$ is a unit vector parallel to the magnetization, $ \mathcal{\bm J}=( \mathcal{J}_x,\mathcal{J}_y,\mathcal{J}_z)$ is the $4 \times 4$ angular momentum matrix, and we assume the magnitude of the exchange (Zeeman) field to be $\Delta_{ex}=0.025 ~\mathrm{eV}$. 

\begin{widetext}
From the $\vec{k}\cdot\vec{p}$ Hamiltonian (Eq.~\eqref{eq:kp}) we can construct an equivalent WSM tight binding (TB) Hamiltonian with next nearest neighbor couplings plus a Zeeman exchange term  (Eq.~\eqref{eq:zeeman}) for a cubic lattice at low energy, as
\begin{equation}
\begin{aligned}
H &= \sum_{\mathbf{r}} c^\dagger_{\mathbf{r}}\,\epsilon_0\, c_{\mathbf{r}} 
+ \sum_{\mathbf{r}}\Big[
c^\dagger_{\mathbf{r}}\,T_x\, c_{\mathbf{r}+ a \hat{x}} 
+ c^\dagger_{\mathbf{r}}\,T_y\, c_{\mathbf{r}+ a \hat{y}}  + c^\dagger_{\mathbf{r}}\,T_z\, c_{\mathbf{r}+ c \hat{z}} 
+ \text{h.c.} \Big] \\
&+ \sum_{\mathbf{r}}\Big[
c^\dagger_{\mathbf{r}}\,T_{xy}\, c_{\mathbf{r}+ a \hat{x}+a \hat{y}}
+ c^\dagger_{\mathbf{r}}\,T_{xz}\, c_{\mathbf{r}+ a \hat{x}+c \hat{z}}+ c^\dagger_{\mathbf{r}}\,T_{yz}\, c_{\mathbf{r}+ a \hat{y}+c \hat{z}} 
+ \text{h.c.} \Big] \\
&- \sum_{\mathbf{r}}\Big[
c^\dagger_{\mathbf{r}}\,T_{xy}\, c_{\mathbf{r}+ a \hat{x}-a \hat{y}}
+ c^\dagger_{\mathbf{r}}\,T_{xz}\, c_{\mathbf{r}+ a \hat{x}-c \hat{z}} + c^\dagger_{\mathbf{r}}\,T_{yz}\, c_{\mathbf{r}+ a \hat{y}-c \hat{z}} 
+ \text{h.c.} \Big] \\
\end{aligned}
\end{equation}
where $\hat{x}$, $\hat{y}$, and $\hat{z}$ are unit vectors pointing in the x, y, and z directions, while $c^\dagger_{\mathbf{r}}$ and $c_{\mathbf{r}})$ are fermionic creation and annihilation operators respectively. The onsite energy $\epsilon_{0}$ and couplings $T_{i}$s are
\begin{equation}
\begin{cases}
    \epsilon_0=(a_0+\frac{4a_1}{a^2}+\frac{2a_2}{c^2})\,I_{4\times 4} + \bigl(\frac{4 c_5}{a^2}+\frac{2 c_3}{c^2}-c_3 m^2\bigr) \,\Gamma^5+\Delta_{ex} \hat{n} \cdot  \mathcal{\bm J} \\
    T_x=-\frac{a_1}{a^2} I_{4\times 4}- \frac{c_4}{ a^2} \Gamma^4-\frac{ c_5}{a^2} \Gamma^5-\frac{\alpha_1 \mathrm{i}}{2 a} \Gamma^{15}-\frac{\alpha_2 \mathrm{i}}{2 a} \Gamma^{23}-\frac{\alpha_3 \mathrm{i}}{2 a} \Gamma^{14}\\
    T_y=-\frac{a_1}{a^2} I_{4\times 4}+ \frac{c_4}{ a^2} \Gamma^4-\frac{ c_5}{a^2} \Gamma^5-\frac{\alpha_1 \mathrm{i}}{2 a} \Gamma^{25}-\frac{\alpha_2 \mathrm{i}}{2 a} \Gamma^{13}+\frac{\alpha_3 \mathrm{i}}{2 a} \Gamma^{24}\\
    T_z=-\frac{ a_2}{c^2} I_{4\times 4} -\frac{c_3}{c^2}\Gamma^5-\frac{v\mathrm{i}}{2c}\Gamma^{35} \\
    T_{xy}=-\frac{c_2}{4a^2}\Gamma^3\\
    T_{yz}=-\frac{c_1}{4ac}\Gamma^1  \\
    T_{xz}=-\frac{c_1}{4ac} \Gamma^2 \\
\end{cases}
\end{equation}
\end{widetext}

\begin{figure*}
\includegraphics[width=\textwidth]{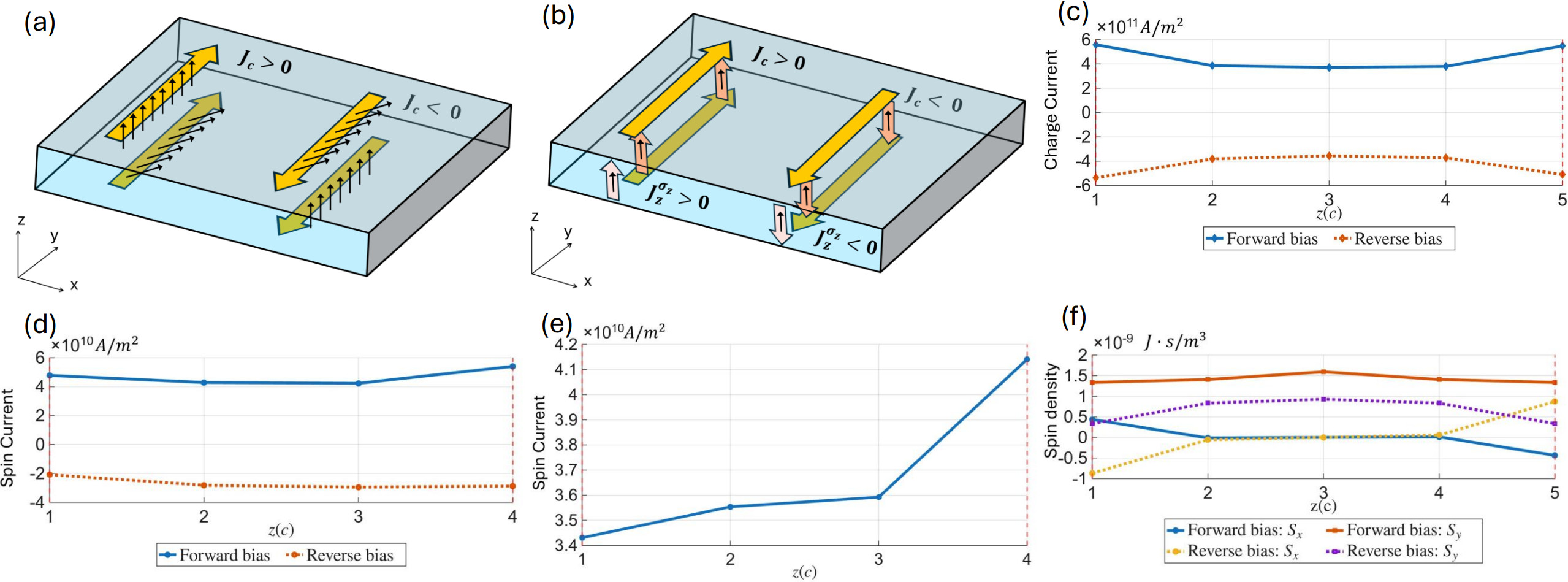}
\caption{\label{fig:WSM} Microscopic features of quantum transport in the Weyl semimetal (WSM) slab of the PIM device during writing mode with in-plane $+\hat{y}$ direction Zeeman field under forward and reverse bias. Under bias voltage, a charge current $J_{y}^{c}$ flows along the $+\hat{y}$ direction within a WSM slab that is infinite in the $x$-direction and biased along $\hat{y}$. Although the system is not translationally invariant in the $y$-direction, the charge current, spin current, and spin density distributions are approximately constant with respect to $y$. Therefore, only depth-dependent ($z$-direction) profiles are shown in the following plots (c)-(f). 
        (a) Inverse spin Galvanic effect (iSGE) mechanism: The spin densities at the top and bottom surfaces (black arrows), have opposite sign for $S_x$ component and the same sign for $S_y$ component, they change sign upon reversing bias. 
        (b) Spin Hall effect (SHE) mechanism: A transverse spin current $J_{z}^{\sigma_{z}}$ (pink arrows), carrying out-of-plane spin angular momentum $+\sigma_z$ (black arrows), flows toward the top surface. 
        There are also non-zero in-plane spin densities $S_x$ and $S_y$ inside the WSM, they change sign upon reversing bias.       
        (c) Depth profile of the charge current density $J_{y}^{c}$ along the $z$-axis, with position measured in units of the lattice constant $c$. 
        (d) Transverse spin current $J_{z}^{\sigma_z}$ carrying $\sigma_z$, plotted between adjacent lattice layers ($i \to i+1$) along $+\hat{z}$-direction. 
        (e) Nonequilibrium component of the transverse spin current that carries out-of-plane spin angular momentum, defined as the antisymmetric part of $J_{z}^{\sigma_z}(V)$ under voltage reversal, i.e., $J^{\sigma_z,\mathrm{neq}}_z(V)=(J_{z}^{\sigma_z}(V)-J_{z}^{\sigma_z}(-V))/2$.
        (f) In-plane spin densities $S_x$ and $S_y$, versus $z$-position under both forward and reverse bias.  }
\end{figure*}

\subsection{\label{subsec:WSMcurrent}Quantum transport in the Weyl semimetal layer}

\begin{widetext}
The nonequilibrium Green's function formalism\cite{datta2018lessons,ghosh2016nanoelectronics,ghosh2023} allows us to calculate the charge and spin currents in a material, by imposing open nonequilibrium boundary conditions to the quantum states. We calculate the currents 
inside the WSM using the model from Ref. \citenum{ruan2016ideal}, described in Sec.~\ref{sec:WSMmodel}, with an exchange (Zeeman) field along varying orientations. We calculate the currents in 2-D, with the transverse coordinate Fourier transformed to generate $\vec{k}_\perp=k_x$ dependent Hamiltonians as below 

\begin{equation}
   \begin{aligned}
&    H_{2D}(k_x)=\sum_{\mathbf{r}}c^\dagger_{\mathbf{r},k_x}\,(\epsilon_0+T_x e^{i k_x a}+T_x^{\dagger}e^{-ik_xa})\, c_{\mathbf{r},k_x} +\Big[
 c^\dagger_{\mathbf{r},k_x}\,T_y\, c_{\mathbf{r}+ a \hat{y},k_x}  + c^\dagger_{\mathbf{r},k_x}\,T_z\, c_{\mathbf{r}+ c \hat{z},k_x}  
+ \text{h.c.} \Big] \\
&+ \sum_{\mathbf{r}}\Big[ c^\dagger_{\mathbf{r},k_x}\,T_{yz}\, c_{\mathbf{r}+ a \hat{y}+c \hat{z},k_x}- c^\dagger_{\mathbf{r},k_x}\,T_{yz}\, c_{\mathbf{r}+ a \hat{y}-c \hat{z},k_x}
+
 c^\dagger_{\mathbf{r},k_x}\,T_{xy}e^{ik_x a}\, c_{\mathbf{r}+ a \hat{y},k_x}-
 c^\dagger_{\mathbf{r},k_x}\,T_{xy}e^{ik_x a}\, c_{\mathbf{r}-a\hat{y},k_x}
+ \text{h.c.} \Big] \\
&+ \sum_{\mathbf{r}}\Big[
 c^\dagger_{\mathbf{r},k_x}\,T_{xz}e^{ik_x a}\, c_{\mathbf{r}+ c \hat{z},k_x}-
 c^\dagger_{\mathbf{r},k_x}\,T_{xz}e^{ik_x a}\, c_{\mathbf{r}-c\hat{z},k_x}
+ \text{h.c.} \Big] \\
   \end{aligned}
\end{equation}
\end{widetext}

In both cases, we attach semi-infinite leads described by self-energies $\Sigma_{\textbf{L/R}}^{R/A/n/m}$ calculated from the recursive surface Green's function \cite{lee1981simple,lee1981simple2,sancho1985highly}. While the contacts are expected to be bulk metals, we use open-boundary conditions corresponding to an extended WSM contact to avoid spurious reflections that are normally taken care of by the numerous contact metallic modes and their internal incoherent scattering events. Different electrochemical potentials at room temperature ($T=300K$) are then attached to the left and right of the device. Upon ignoring spatial impurity scattering, phonons and strong Coulomb interactions, the  Green's functions with different transverse momentum ($k_x$) and energies decouple. The lesser Green's functions can then be solved by the Dyson and Keldysh equation(i.e., $G^{R/A}(E,k_x)= [E\pm i \eta -H_{2D}(k_x)-\Sigma^{R/A}_{L}(E,k_x)-\Sigma^{R/A}_{R}(E,k_x)]^{-1}$ and $G^{n}(E,k_x)=G^{R}\Sigma^{n}G^{A}$) numerically. We can calculate the spin densities from the lesser Green's function by
\begin{equation}
  S_{i}=\frac{\hbar}{2 V}\int\frac{dk}{2 \pi}\int \frac{dE}{2\pi} \operatorname{Tr} \left(  \mathcal{J}_s G_{ii}^{n} \right)
    \label{eq:Spin density}
\end{equation}

\begin{widetext}
The  expressions for charge and spin current densities are \cite{meir1992landauer,datta1997electronic}
\begin{equation}
J_{ij}^{c} = \frac{q}{\hbar A}\int\frac{dk}{2 \pi}\int \frac{dE}{2\pi} \operatorname{Im} \left\{ \operatorname{Tr} \left[ H_{2D,ji}\, G^{n}_{ij} - G^{n}_{ji}\, H_{2D,ij} \right] \right\}
\label{eq:Jc}
\end{equation}
\begin{equation}
J_{ij}^{\sigma_s} = \frac{q}{\hbar A}\int \frac{dk}{2 \pi}\int \frac{dE}{2\pi} \operatorname{Im} \left\{ \operatorname{Tr} \left[ \mathcal{J}_s \left( H_{2D,ji}\, G^{n}_{ij} - G^{n}_{ji}\, H_{2D,ij} \right) \right] \right\}
\label{eq:Js}
\end{equation}
Here, $i$ and $j$ are lattice-site indices in the tight-binding model. $V$ is the volume of a single unit cell. We measure both charge current and spin current in units of $\mathrm{A/m^2}$. $\mathcal{J}_s$ is the spin matrix for a spin-$\frac{3}{2}$ system, $A$ is the unit-cell cross-sectional area perpendicular to the current direction, while $q$ and $\hbar$ are the elementary charge and reduced Planck's constant, respectively.   
\end{widetext}

Because the four-band low-energy model includes couplings up to next-nearest neighbors in the $y-z$ plane, the nonzero charge and spin bond--current contributions arise only from pairs of sites $i$ and $j$ with relative displacement
$\mathbf r_j-\mathbf r_i \in \{\pm a\,\hat{\mathbf y},\; \pm c\,\hat{\mathbf z},\; \pm a\,\hat{\mathbf y} \pm c\,\hat{\mathbf z}\}$
and these bonds connect modes with the same transverse momentum $k_x$.  
Since the bias voltage drops along the $y$ direction, both the total net charge current and longitudinal spin current flow in the $\pm \hat{y}$ direction. $J_{y}^{c/\sigma_s}$ is thus related to the currents between bonds by $J_{y}^{c/\sigma_s}=J_{i,i+a\hat{y}}^{c/\sigma_s}+\cos\theta_{yz} J_{i,i+a\hat{y}+c\hat{z}}^{c/\sigma_s}+\cos\theta_{yz} J_{i,i+a\hat{y}-c\hat{z}}^{c/\sigma_s}, ~~~(\cos \theta_{yz}=\frac{a}{\sqrt{a^2+c^2}})$. We are mostly interested in spin currents flowing towards or away from the surface of the Weyl semimetal, so we define the spin current flowing along the $+\hat{z}$ direction as the transverse spin current $J_{z}^{\sigma_s}$. The transverse spin currents are related to the currents between bonds by $J_{z}^{c/\sigma_s}=J_{i,i+c\hat{z}}^{c/\sigma_s}+\cos\theta_{zy} J_{i,i+a\hat{y}+c\hat{z}}^{c/\sigma_s}+\cos\theta_{zy} J_{i,i-a\hat{y}+c\hat{z}}^{c/\sigma_s}~~(\cos \theta_{zy}=\frac{c}{\sqrt{a^2+c^2}})$. 

The spin current induced by the spin Hall effect can be expressed as
\begin{equation}
J_{z}^{\vec{\sigma}} = 
    \alpha_{\text{trans}} B_z \, (\vec{\cal{E}}_{\parallel} \cdot \vec{\sigma})
    + \left(
        \alpha_{\text{planar}} \, \vec{B}_{\parallel} \cdot \vec{\cal{E}}_{\parallel}
        + \alpha_{\text{long}} \, B_z {\cal{E}}_z
    \right) (\vec{\sigma} \cdot \hat{z}).\label{eq:spincurrentfull}
\end{equation}
Here $\vec{\cal{E}}=\vec{\cal{E}}_{\parallel}+{\cal{E}}_z\hat{z}$ is the electric field, and $\vec{B}=\vec{B}_{\parallel}+B_z\hat{z}$ is the effective exchange (Zeeman) term responsible for modulating the symmetry of the crystal. More specifically, $\vec{B}_{\parallel}= (B_x, B_y,0)$ and $\vec{\cal{E}}_{\parallel}= ({\cal{E}}_x, {\cal{E}}_y,0)$ are the in-plane electric and exchange (Zeeman) fields, and $\alpha_{\textbf{trans}}$, $\alpha_{\textbf{planar}}$, as well as $\alpha_{\textbf{long}}$ describe the transverse (in-plane $\vec{\cal{ E}}_{\parallel}$ and out-of-plane $B_z$) coupling, planar (in-plane $ \vec{\cal{ E}}_{\parallel}$ and in-plane $\vec B_{\parallel}$) coupling, and longitudinal (out-of-plane ${\cal{E}}_z$ and out-of-plane $B_z$) coupling. This formula is set by the group theoretical symmetry $D_{2d}$ of the crystal. The transverse spin current induced by planar magnetoelectric coupling $\alpha_{\textbf{planar}} {\cal{E}}_y B_y$ is the key term in building the device discussed in this paper.

 For conventional topological insulator materials (TIs) such as $\mathrm{Bi_{2}Se_{3}}$\cite{vakili2022low}, the spin Hall effect (SHE) term is negligible, and an out-of-plane exchange (Zeeman) field gaps the surface state, making TIs insulating. For the Weyl semimetal discussed in this paper, the bulk term is controlled by the gate bias, as well as the orientations and magnitude of the exchange (Zeeman) field, as indicated by Eq.~\eqref{eq:spincurrentfull}.
 

The spin current induced by the inverse spin Galvanic effect is more complicated. For conventional topological insulators and Weyl semimetals\cite{johansson2018edelstein,limtragool2022large}, the in-plane components mostly perpendicular to the electric field are controlled by the gate bias
\begin{equation}
    J_{z,{\mathrm{iSGE}}}^{\vec{\sigma}} = \lambda_{\perp} \left({\vec{\cal{E}}}\times\hat{z}\right)\cdot\vec{\sigma}+\lambda_{\parallel}{\vec{\cal{E}}}\cdot \vec{\sigma}
\end{equation}
where $\lambda_{\parallel}$ and $\lambda_{\perp}$ are parameters that depend on the details of the device and material system. For conventional topological insulators, we mostly have iSGE current with polarization perpendicular to the electric field, i.e., $\lambda_{\perp}\neq0,\lambda_{\parallel}=0$. This current induced in-plane spin polarization acts as the dominant source of spin–orbit torque (SOT), enabling magnetization switching between the $-\hat{x}$ and $+\hat{x}$ directions\cite{manchon2019current,mellnik2014spin,fischer2016spin,han2017room,vakili2022low}. By contrast, for the Weyl semimetal discussed in this paper, the iSGE current has polarization both perpendicular to and parallel to the electric field. The iSGE current is also controlled by both the exchange (Zeeman) field and the bias voltage. That is, both $\lambda_{\perp}$ and $\lambda_{\parallel}$ are functions of the electric field and the exchange (Zeeman) field. 
\subsection{\label{sec:LLG}LLG simulation of magnetization dynamics}
 The magnetization dynamics of the free and selector magnets can be described by the Landau-Lifshitz-Gilbert (LLG) equation
\begin{align}\label{eq:LLG}
\frac{1 + \alpha^2}{\gamma} \cdot \frac{\partial \bm{m}}{\partial t} =
- \mu_0 \cdot (\bm{m} \times \bm{H}_{\mathrm{eff}}) \\ \nonumber
- \alpha \mu_0 \cdot \bm{m} \times (\bm{m} \times \bm{H}_{\mathrm{eff}})\\ \nonumber
- \frac{\hbar}{2q} \cdot \frac{J_s}{M_{s} t_{FM}} \cdot \bm{m} \times (\bm{m} \times \bm{\sigma}_p).\\   \nonumber
\end{align}
Here $\bm m=\bm M/M_s$ is the unit magnetization (with magnetization $\bm M$ and saturation magnetization $M_s$), $\alpha$ is the Gilbert damping, $\gamma$ is the gyromagnetic ratio, and $\mu_0$ is the vacuum permeability. The damping-like torque term is set by $\frac{\hbar}{2q} \cdot \frac{J_s}{M_{s} t_{FM}} \cdot \bm{m} \times (\bm{m} \times \bm{\sigma}_p)$, where $J_s$ is the magnitude of the spin current density injected by the Weyl semimetal (unit: $\mathrm{A/m^2}$), $\bm{\sigma}_p$ is a unit vector along the spin polarization of the spin current and $t_{\mathrm{FM}}$ is the ferromagnet thickness. 

The effective magnetic field consists of four parts $\bm H_{eff}=\bm H_{anis}+\bm H_{demag}+\bm H_{stress}+\bm H_{th}$, comprising the anisotropy, demagnetizing, stress-induced, and thermal contributions.
For the free magnet, $\bm H_{anis}=\frac{2 K_u}{\mu_0 M_s}(\hat{z} \cdot \bm m)\hat{z}$ is the effective anisotropy field, $\bm H_{demag}=-M_s (\hat{z} \cdot \bm m) \hat{z}$ is the demagnetization field, the stress-induced field is $\bm 0$ and the thermal field is Gaussian distributed with zero mean and standard deviation $\mathrm{SD} = \sqrt{\frac{2\alpha k_B T}{\mu_0^2 \gamma M_s V \Delta t}}$, where $V$ is the volume, $T$ is room temperature, $k_B$ is the Boltzmann constant and $\Delta t$  is the simulation time step\cite{xie2023anatomy}. In Fig.~\ref{fig:magnet} (d), we start from $\bm m=(0,0,1)$ and turn on the damping-like torque term using spin current with polarization $-\sigma_z$ and magnitude $J_{z}^{\sigma_z,\text{write "0"}}=3.18\times 10^{10}~\mathrm{A/m^2}$ for $5~\mathrm{ns}$. Similarly, in Fig.~\ref{fig:magnet}.
For the selector magnet, the anisotropic field, demagnetization field, and thermal field have the same form but different parameters, as shown in Table.~\ref{table:Material parameters}. We have a stress-induced field produced by the piezo $\bm H_{stress}= \frac{3\lambda_s \sigma}{\mu_0 M_{s}} (\hat{y} \cdot \bm m) \hat{y}$, where $\lambda_s$ is the magnetostrictive coefficient of the gating magnet, and $ \sigma=Y \epsilon$ is the stress generated by the electrical strain induced by the piezoelectric\cite{abeed2018magneto,winters2019reliability,mishra2023strain}. Here $Y$ denotes Young's modulus and $\epsilon$ denotes the strain generated by PZT. The damping-like torque is off in the selector simulation. For Fig.~\ref{fig:magnet} (e), we start from $\bm m=(0,0,1)$ and turn on the stress-induced field. For Fig.~\ref{fig:magnet} (f), we start from $\bm m=(0,1,0)$ without turning on the stress-induced field. Material parameters for the piezoelectric layer are adopted directly from Ref.~\citenum{morshed2025strained}, which provides detailed calculations and analyses of the piezoelectric response, including stress, strain, and related material properties (see also the references therein). Parameters for the selector and free magnets are also based on Ref.~\citenum{morshed2025strained} and the references therein, with some adjustments to incorporate the characteristics of the Weyl semimetal system considered in this work.

\begin{table}[h!]
\centering
\caption{\label{table:Material parameters}Material parameters\cite{morshed2025strained}}
\begin{tabular}{|c|c|}
\hline
\textbf{Dimension} & \\
\hline
Length $L$ & 33.5 nm \\
Width $W$ & 33.5 nm \\
\hline
\multicolumn{2}{|c|}{\textbf{Selector Magnet: $\mathrm{TbCo}$}} \\
\hline
Thickness $t_{FM}$ & 12.5 nm \\
Uniaxial anisotropy $K_{u}$ & $3 \times 10^5$ $\mathrm{J/m^3}$\cite{abeed2018magneto,aviles2025perpendicularly}  \\
Saturation magnetization $M_{s}$ & $6 \times 10^5$ $\mathrm{A/m}$ \cite{aviles2025perpendicularly,suzuki2023thickness,thorarinsdottir2023competing}\\
Damping $\alpha$ & 0.8 \\
Magnetostrictive coefficient $\lambda_s$ & $400 \times 10^{-6}$ \cite{betz1999magnetic} \\
Young’s modulus $Y$ & $100 \times 10^9$ $\mathrm{Pa}$ \cite{yoshino1984perpendicular} \\
\hline
\multicolumn{2}{|c|}{\textbf{MTJ Free Layer: $\mathrm{TbCo}$}} \\
\hline
Thickness $t_{FM}$ & 1.5 nm \\
Uniaxial anisotropy $K_{u}$ & $3 \times 10^5$ $\mathrm{J/m^3}$\cite{abeed2018magneto,aviles2025perpendicularly} \\
Saturation magnetization $M_{s}$ & $8 \times 10^4$ $\mathrm{A/m}$ \cite{honda2003magnetization,aviles2025perpendicularly,suzuki2023thickness,finley2016spin}\\
Damping $\alpha$ & 0.01 \\
\hline
\multicolumn{2}{c}{\textbf{Weyl semimetal: $\mathrm{CuTlTe_2}$}}\\
\colrule
Thickness $t_{\mathrm{WSM}}$ & 6.4 nm (5 layers) \\
Surface spin Hall angle $\theta_{\mathrm{sh,surface}}$ &
  \begin{tabular}[t]{@{}l@{}}
  $\sim 0.0463$ (writing "0")\\
  $\sim 0.0782$ (writing "1")
  \end{tabular} \\
Total spin Hall angle $\theta_{\mathrm{sh,total}}$ &
  \begin{tabular}[t]{@{}l@{}}
  $\sim 0.0115$ (writing "0")\\
  $\sim 0.0188$ (writing "1")
  \end{tabular} \\
\hline
\multicolumn{2}{|c|}{\textbf{Piezoelectric: $\mathrm{PZT}$}} \\
\hline
Thickness $t_{\text{piezo}}$ & 100 nm \\
Piezoelectric constant $d_{31}$ & $1.8 \times 10^{-10}$ m/V \cite{roy2011hybrid} \\
Max. strain $\epsilon$ & 1000 ppm (0.1\%) \cite{roy2012energy} \\
Relative dielectric constant $\epsilon_r$ & 1000 \cite{roy2011hybrid} \\
\hline
\end{tabular}
\end{table}

\section{\label{sec:Results} Results and discussion}

\subsection{\label{sec:Charge‑to‑Spin Conversion Efficiency}Charge‑to‑Spin Conversion Efficiency of WSM}


As indicated by Eqs.~\eqref{eq:spin current1} and \eqref{eq:spincurrentfull}, to generate spin currents carrying out-of-plane angular momentum, we can apply an exchange (Zeeman) field with a finite component along the $\hat{y}$ direction. This exchange (Zeeman) field explicitly breaks the two rotational symmetries $C_{2x}$ and $C_{2z}$, two mirror symmetries ($M_{xy}$ and $M_{−xy}$), and improper rotations ($S_{4z}^{+}$ and $S_{4z}^{-}$), while all crystal symmetry operations are illustrated in Fig.~\ref{fig:D2d}, along with the time-reversal symmetry $\mathcal{T}$. Compared to Figs.~\ref{fig:Band and Symmetries}~(b),~~(e), which preserve all $D_{2d}$ point group crystal symmetries, as reflected by the surface state (b) and bulk state (e), the Fermi arc of the Weyl semimetal with an exchange (Zeeman) field in the $+\hat{y}$ direction, shown in Figs.~\ref{fig:Band and Symmetries}~(c),~(f), reflects the breaking of rotational symmetry $C_{2z}$ and mirror symmetries for both the surface state (c) and bulk state (f). At this point, the transverse spin current in $\hat{z}$ direction carrying out-of-plane spin angular momentum induced by charge current in the  $\hat{y}$ direction does not vanish due to symmetry. In contrast, the exchange (Zeeman) field in the $\hat{z}$ direction does not break the $C_{2z}$ rotational symmetries, and the transverse spin current in $\hat{z}$ direction carrying out-of-plane spin angular momentum induced by the charge current in the $\hat{y}$ direction vanishes due to symmetry. Alternatively, the exchange (Zeeman) field–dependent spin Hall conductance satisfies $\sigma^{z}_{zy}(B_z \hat{z}) = 0$, whereas $\sigma^{z}_{zy}(B_y \hat{y}) \neq 0$.

Fig.~\ref{fig:WSM} shows theoretical predictions of the quantum transport properties of a Weyl semimetal in the presence of an exchange (Zeeman) field oriented along the $+\hat{y}$ direction in numerical TB-NEGF calculations. Unlike conventional topological insulators such as $\mathrm{Bi_{2}Se_{3}}$, whose spin-orbit torque comes from the iSGE with in-plane polarized spin perpendicular to both the charge current direction, Weyl semimetals with exchange (Zeeman) field have both iSGE and SHE that contribute to spin-orbit torque, as shown in Fig.~\ref{fig:WSM} (a) and (b). The iSGE originates from the spin Edelstein effect in the Weyl semimetal, where a charge current induces nonequilibrium spin polarization on its surface. 

\begin{figure*}
\includegraphics[width=\textwidth]{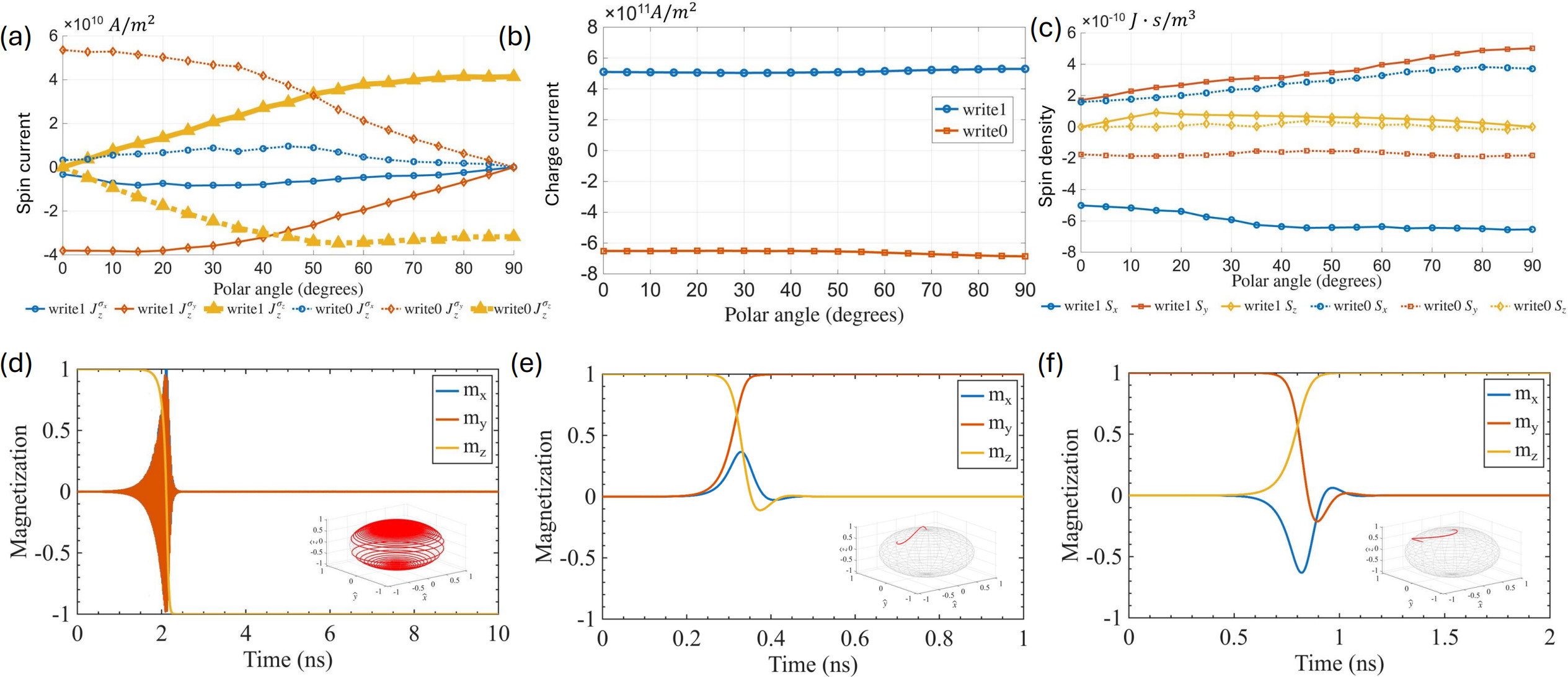}
\caption{\label{fig:magnet} Dynamical properties for WSM layer, free magnet layer and selector magnet layer:
(a) Relation between nonequilibrium spin current and the polar angle of the Zeeman field in the $y-z$ plane relative to the $z$-axis. The azimuthal angle is $90^{\circ}$, while the polar angle runs from $0^{\circ}$ to $90^{\circ}$. Provided by the selector magnet, the exchange Zeeman coupling between the selector magnet and the WSM is $\Delta_{ex} = 0.025~\mathrm{eV}$ and the bias voltage is $V=-0.1~\mathrm{V}$ for writing "0" and $V=0.1~\mathrm{V}$ for writing "1".
(b) Relation between charge current and the polar angle of Zeeman field in the $y-z$ plane relative to the $z$-axis.
(c) Relation between nonequilibrium spin densities on the surface layer and the polar angle of Zeeman field in the $y-z$ plane relative to the $z$-axis.
(d) Writing a "0": switching the magnetization of the magnet from $+\hat{z}$ to $-\hat{z}$. 
(e) Changing from storage mode to writing mode by applying gate voltage to piezo and rotating the magnet from $+\hat{z}$ to $+\hat{y}$
(f) Changing from writing mode to storage mode, by not applying gate voltage to the piezo and rotating the magnet from $+\hat{y}$ to $+\hat{z}$. 
}
\end{figure*}

For Weyl semimetals with an exchange (Zeeman) field, the charge current mainly induces in-plane spin polarization, delivering torque that can rotate the magnet only in-plane if we do not consider the SHE. In addition to the spin-orbit torque induced by iSGE, Weyl semimetals may also exhibit the spin Hall effect (SHE), which may generate a transverse spin current that carries out-of-plane spin angular momentum exerted on the nearby magnet, thereby rotating the magnet vertically. However, the SHE is controlled by both the symmetry-breaking exchange (Zeeman) field and the bias voltage. As shown in Eq.~\eqref{eq:spincurrentfull} and Fig.~\ref{fig:magnet} (a), for Weyl semimetals with an exchange (Zeeman) field oriented along the $\pm \hat{z}$ direction, the transverse spin current only delivers in-plane $\sigma_x$ and $\sigma_y$ spin angular momentum. In contrast, when the exchange (Zeeman) field is oriented along $\pm \hat{y}$, the transverse spin current carries out-of-plane $\sigma_z$ spin angular momentum, i.e., $\pm J_{z}^{\sigma_z}$, the sign of which reverses with the field orientation and also reverses with the bias voltage.

In this theoretical prediction, we assumed an infinitely long Weyl semimetal slab that is infinite in the $x$ direction, applied a finite bias voltage across the $\hat{y}$ direction, and calculated the charge current $J_{y}^{c}$ flowing in the $\hat{y}$ direction, the transverse spin current $J_{z}^{\sigma_s}$, and the nonequilibrium spin polarization in the sample (Eqs.~\ref{eq:Jc}, ~\ref{eq:Js},~\ref{eq:Spin density}). 
The charge current $J_{y}^{c}(V)$  under bias voltage $V$ flows along the $\hat{y}$ direction and changes sign upon reversing the bias voltage, as shown in Fig.~\ref{fig:WSM} (c). 

When applying the exchange (Zeeman) field in the $+\hat{y}$ direction, we found that the transverse spin current that carries in-plane spin angular momentum vanishes, i.e., $J_{z}^{\sigma_x}=J_{z}^{\sigma_y}=0$. However, as shown in Fig.~\ref{fig:WSM} (d), a finite transverse spin current with out-of-plane polarization is not identical upon reversing the bias voltage, i.e., $J_{z}^{\sigma_z}(V)\neq J_{z}^{\sigma_z}(-V)$, because it contains both equilibrium (even-in-$V$) and nonequilibrium (odd-in-$V$) contributions. To extract the nonzero nonequilibrium contributions to $J_{z}^{\sigma_z}$ , one can calculate the antisymmetric part of $J_{z}^{\sigma_z}(V)$ under voltage reversal, i.e., $J^{\sigma_z,\mathrm{neq}}_{z}(V)=(J_{z}^{\sigma_z}(V)-J_{z}^{\sigma_z}(-V))/2$. The spin Hall angle can be defined as
\begin{equation}
    \theta_{SH}=\frac{J_{z}^{\sigma_z}(V)-J_{z}^{\sigma_z}(-V)}{J_{y}^{c}(V)-J_{y}^{c}(-V)}
\end{equation}

For the surface spin Hall angle, we should divide the transverse spin current flowing towards the surface by the charge current flowing in the surface layer. For the total spin Hall angle, we should divide the transverse spin current flowing towards the surface by the total charge current across a section parallel to the $x-z$ plane. The spin Hall angle is listed in Tab.~\ref{table:Material parameters}. The distribution of nonequilibrium contributions to $J_{z}^{\sigma_z}$ is shown in Fig.~\ref{fig:WSM} (e). Also, we plot the spin density $S_i$ distribution inside the Weyl semimetal in Fig.~\ref{fig:WSM} (f). For the exchange (Zeeman) field in the $+\hat{y}$ direction, the spin density has only an in-plane $S_x$ and $S_y$ part. The spin density distribution contains both the Zeeman field induced and current induced parts. We can also extract the current induced spin density by subtracting the spin density under forward bias voltage $S_i(V)$ from the spin density under reverse bias voltage  $S_i(V)$, i.e., $S_{i,neq}(V)=(S_i(V)-S_i(-V))/2$.

The magnitude of the transverse spin current $J_{z}^{z,neq}$ can be controlled by both the bias voltage and the direction and magnitude of the exchange (Zeeman) field. Fig.~\ref{fig:magnet} (a) shows the relation between the sign and magnitude of the nonequilibrium transverse spin current $J_{z}^{\sigma_s,neq}$ under opposite drain bias voltages ($V_{drain}=-0.1V$ for writing "0" and $V_{drain}=0.1V$ for writing "1"), with the exchange (Zeeman) field rotating in the $y-z$ plane from the $+\hat{z}$ direction with a polar angle equal to $0^{\degree}$ to the $+\hat{y}$ direction with a polar angle equal to $90^{\degree}$. The results agree qualitatively with Eq.~\eqref{eq:spincurrentfull} from symmetry analysis. We are mostly interested in the transverse spin current that carries out-of-plane spin angular momentum $J_{z}^{\sigma_z}$, since it can rotate a magnet vertically. This transverse spin current $J_{z}^{\sigma_z}$ vanishes when the exchange (Zeeman) field is pointing along the$+\hat{z}$ direction and becomes larger as the polar angle approaches $90^\circ$. For polar angles close to $90^\circ$, the current may slightly decrease, but the overall trend remains consistent with theoretical expectations. When the exchange (Zeeman) field is pointing along the $+\hat{y}$ direction, the magnitude of the transverse spin current $J_{z}^{\sigma_z,\mathrm{neq}}$ reaches a large value that can rotate a magnet vertically. However, the transverse spin currents with the exchange (Zeeman) field in the $\hat{y}$ direction for writing "0" and "1" are different. The spin current generated by the Weyl semimetal for write-in "0" is $J_z^{\sigma_z,\text{write "0"}}=3.18 \times 10^{10}~\mathrm{A/m^2}$, and the spin current for write-in "1" is $J_{z}^{\sigma_z,\text{write "1"}}=4.14 \times 10^{10}~\mathrm{A/m^2}$. In the simulation, we set the grounded source chemical potential equal to the energy of the Weyl point, i.e., $\mu_{Source}=E_{Weyl}$, and tuned the drain chemical potential to $\mu_{Drain,\text{write "0"}}=\mu_{Source}-V$ and $\mu_{Drain,\text{write "1"}}=\mu_{Source}+V$. The asymmetry of the spin current arises from the asymmetry of the spectrum near the Weyl point with respect to a Fermi surface penetrating the Weyl points, as shown in Fig.~\ref{fig:Band and Symmetries} (d).


 \begin{figure*}
\includegraphics[width=\textwidth]{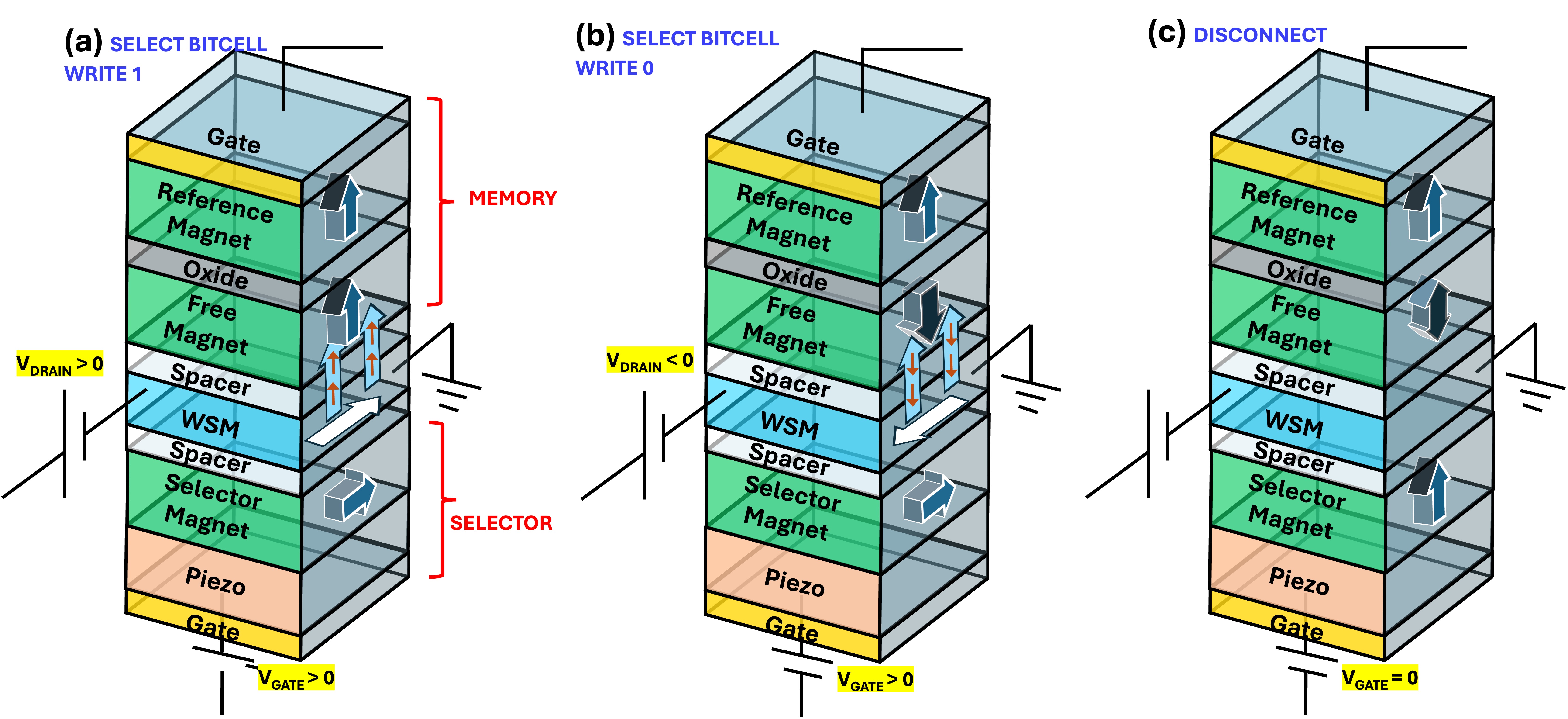}
\includegraphics[width=0.6\textwidth]{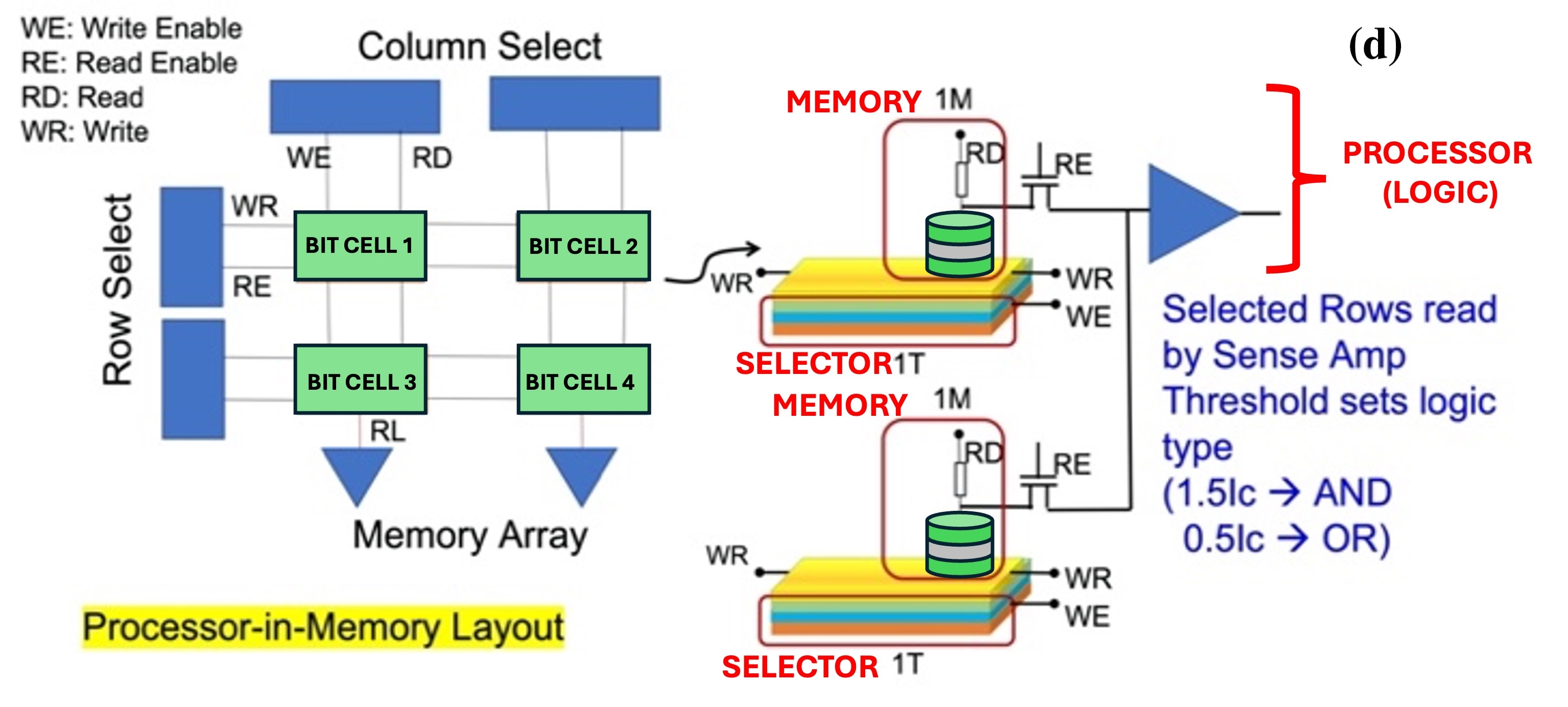}
\caption{\label{fig:Device} Device geometry and three modes of the device:  (a) Writing "1": When the drain voltage $V_{drain}>0$ and the gate voltage $V_{gate}>0$ such that the magnetization of the selector magnet is in $+\hat{y}$ direction, the charge current (white arrow) flows from drain to source and the WSM generates transverse spin currents (light blue arrow) that carries $+\sigma_z$ spin angular momentum (orange arrow) flowing towards the storage magnet. The magnetization of the storage magnet can be rotated to $+\hat{z}$ direction and write "1". (b) Writing "0": When the drain voltage $V_{drain}<0$ and the gate voltage $V_{gate}>0$ such that the magnetization of the selector magnet is in $+\hat{y}$ direction, the charge current (white arrow) flows from drain to source and the WSM generates transverse spin currents (light blue arrow) that carries $-\sigma_z$ spin angular momentum (orange arrow) flowing towards the storage magnet. The magnetization of the storage magnet can be rotated to $-\hat{z}$ direction and write "0". (c)Storage mode: When no gate voltage is applied to the piezo, i.e., $V_{gate}=0$, the magnetization of the selector magnet is in $\pm \hat{z}$ direction. As a result, the WSM does not generate transverse spin currents that carries out-of-plane spin angular momentum, regardless of what the drain voltage $V_{Drain}$ is and whether there is charge current flowing between source and drain. The storage magnet cannot be rotated and the device is in storage mode. (d) The overall PIM structure, that selects two rows and runs them through a sense amplifier to carry out a bitwise AND or OR logic operation, building thereby one processor locally around a few proximal non-volatile logic elements.}
\end{figure*}
\subsection{\label{sec:Freemagnet Switching Dynamics}Free Magnet Switching Dynamics}
The free magnet is part of a magnetic tunnel junction (MTJ) with an out-of-plane polarized fixed magnet and a tunnel barrier. The binary data encoded in its magnetization along $\pm \hat{z}$ directions can be read through the tunneling magnetoresistance of the MTJ, and can be overwritten by the out-of-plane spin angular momentum transferred from the Weyl semimetal. 
As shown in Fig.~\ref{fig:magnet}(a), the transverse spin current $J_{z}^{\sigma_z}$ depends on the orientation of the exchange (Zeeman) field, delivering $\pm \sigma_{z}$ directed spins when the exchange field orients along $+\hat{y}$ directions, but not when it is oriented along $+\hat{z}$. In contrast, as shown in Fig.~\ref{fig:magnet}(b), the surface charge current densities remain almost unchanged; and, as shown in Fig.~\ref{fig:magnet}(c), the Weyl semimetal mainly exhibits in-plane current induced surface spin polarizations for the spin Edelstein effect. The switching is governed by the out-of-plane spin transfer driven by the spin Hall effect, and the critical spin current associated with the transverse spin current carrying out-of-plane spin angular momentum can be estimated by balancing the spin torque against the magnetic damping and the effective magnetic field due to anisotropy as well as demagnetization:\cite{timopheev2015respective,sun2000spin,sun1999current} 
\begin{equation}
    J_s^{\text{crit}} = \frac{2q}{\hbar} \, \alpha \mu_{0} M_s H_{\text{eff}} \, t_{\text{FM}},
\end{equation}
We design the geometry of the free magnet such that the effective magnetic field $\mathbf{H}_{\text{eff}}$ is aligned along the $\hat{z}$ direction with a magnitude $|\mathbf{H}_{\text{eff}}|=\frac{2 K_u}{\mu_0 M_s}-M_s$. The parameters of the free magnet are listed in Tab.~\ref{table:Material parameters}. In our model, the critical spin current carrying out-of-plane spin angular momentum is estimated to be $J_s^{\text{crit}} =2.70 \times 10^{10}~\mathrm{A/m^2}$, which is smaller than the transverse spin current generated by the Weyl semimetal to write information "0" or "1" when the exchange (Zeeman) field is in $+ y$ directions. This allows semi-deterministic switching of the free magnet in a short time $< 5~\mathrm{ns}$ and with high accuracy, as shown in Fig.~\ref{fig:magnet} (d). In our simulations, we omit the iSGE because its in-plane spin accumulation has little influence on the switching process. Instead, we focus on the SHE, which delivers out-of-plane spin angular momentum parallel to the anisotropy axis and thus primarily governs the magnetization switching. 

\subsection{\label{sec:selector Switching Dynamics}Rotating the selector magnet with voltage gated strain}

Figs.~\ref{fig:Device} (a), (b), and (c) show a selector magnet placed between the Weyl semimetal and a piezoelectric actuator. The selector magnet provides an effective exchange (Zeeman) field with a coupling strength of $\Delta_{ex} = 0.025  ~\mathrm{eV}$ acting on the Weyl semimetal. As discussed in Sec.~\ref{sec:Charge‑to‑Spin Conversion Efficiency} and shown in Fig.~\ref{fig:magnet}(a), the magnetization orientation of the selector magnet determines the orientation of the exchange (Zeeman) field in the Weyl semimetal. Consequently, it modifies the symmetries of Weyl semimetals and controls the ON and OFF states of the transverse spin current that carries the out-of-plane spin angular momentum by Eq.~\eqref{eq:spin current1}. On the other hand, in addition to the uniaxial anisotropy field $\bm H_{anis}$ and the demagnetization field $\bm H_{demag}$, the piezoelectric element attached to the selector magnet also provides an effective stress-induced field $\bm H_{stress}$ aligning in $\hat{y}$, which rotates the magnetization of the selector magnet between the $\pm \hat{z}$ and $\pm \hat{y}$ directions.  The magnetization dynamics are governed by the competition between the uniaxial anisotropy field, the demagnetization field, and the effective stress-induced field. 

Fig.~\ref{fig:magnet}(e) and (f) show that turning the stress-induced field on rotates the selector magnet from the $+\hat{z}$ direction to the $+\hat{y}$ direction in less than $1~\mathrm{ns}$. When the stress-induced field is turned off, the magnetization of the selector magnet returns to the $+\hat{z}$ direction in approximately $1 ~\mathrm{ns}$. The switching of the selector magnet is faster than the switching of the free magnet, preventing the unintentional switching of the free magnet.

\subsection{\label{sec:Device}Design of Process-in-Memory Device}
We now discuss the geometry of the four-terminal SWSM-SOT-MRAM that can integrate the selector and memory in a single bitcell. The SWSM-SOT-MRAM needs to execute three local operations: read, write, and row-column select.

The read unit, the MTJ, consists of a fixed reference magnet along the $+\hat{z}$ direction magnetization and a reading line terminal attached on top; a free storage magnet, whose magnetization can be rotated by a spin current carrying out-of-plane ($\pm \sigma_z$) spin angular momentum from the Weyl semimetal beneath it, which is attached to a drain for readout; and an oxide layer placed between the reference magnet and the storage magnet. The binary digit information, encoded in the magnetization directions along $\pm \hat{z}$ of the storage magnet, can be read from the tunneling current through the MTJ. In storage mode, the magnetization of the selector magnet is placed parallel to the $\pm \hat{z}$ directions, whereupon the symmetry of the Weyl semimetal cancels any  transverse spin current arising from bulk SHE states.

The write mechanism arises from the Weyl semimetal that can generate a transverse spin current with out-of-plane spin angular momentum, and a spacer that can control the interaction strength between the Weyl semimetal and the free storage magnet. The write mode triggers when the magnetization of the selector magnet is placed parallel to the $\pm \hat{y}$ directions, so that the exchange interactions between the selector magnet and the Weyl semimetal break the time-reversal and discrete spatial symmetries of the Weyl semimetal. At positive (negative) drain bias, charge currents flow forward (backward) between the drain and the grounded source, generating transverse spin current that carries $+\sigma_z$ ($-\sigma_{z}$) spin angular momentum. 

Finally, the row-column selection arises from the gate controlled piezo sitting below the selector magnet, with a spacer separating it from the WSM to tune its proximity exchange interaction.  When the piezo is relaxed, it exerts no strain on the selector magnet, whose out-of-plane anisotropy places its magnetization along the $\pm \hat{z}$ direction. Under positive bias gate voltage on the piezo, the shear strain creates an effective magnetic field along y via magnetoelastic coupling, rotating the selector magnet along $\pm \hat{y}$, and bringing it into write mode. 

The magnetodynamics of the selector magnet is shown in Fig.~\ref{fig:magnet}. Note that for the TI the y-degeneracy ($\pm \hat{y}$) of switching the bottom selector magnet is irrelevant, as either in-plane orientation preserves the TI surface states, while both out-of plane orientations $\pm \hat{z}$ gap them. In contrast, for the WSM the selector magnet orientation $+\hat{y}$ vs $-\hat{y}$ sets the exchange field which in turn sets the polarity along $+\hat{z}$ vs $-\hat{z}$ of the spin delivered to the write magnet, only one of which can deliver an anti-damping torque to the write magnet. To fix this, we can either bias the selector magnet towards one of the polarities, such as with a field-assist or a dipole/exchange field from a neighboring magnet, or else flip the drain bias if we get the wrong polarity of the spins delivered.

\section{\label{sec:Data availability}Data availability}
All data (parameter tables, plots, underlying Hamiltonians, and equations) are included in the paper.
\section{\label{sec:Code availability}Code availability}
Codes involved may be available upon request. 
\section{\label{sec:Acknowledgments}Acknowledgments}
We acknowledge useful discussions with Saroj Dash, Joe Poon, Dmytro Pesin, Supriyo Bandyopadhyay, Joe Hagmann, Dan Gopman and Patrick Taylor. This project was funded by NSF IUCRC 1939012 and NSF CISE 2504227 grants. 
\section{\label{sec:Author contributions}Author contributions}
YC did the bulk of the calculations, building extensively on prior software developed by HV and GM. AWG was the project lead, conceptualized the PIM structure. All authors collaborated on the physics, specifically the switching and symmetry breaking mechanisms, the PIM device design and the write-up.
\section{\label{sec:Competing Interests}Competing Interests}
None
\bibliography{submission-to-arxivprb}
\bibliographystyle{apsrev4-2} 

\end{document}